\documentclass[lettersize,journal]{IEEEtran}
\usepackage{cite}
\usepackage{amsmath,amssymb,amsfonts}
\usepackage{booktabs}
\usepackage{makecell}
\usepackage{algorithmic}
\usepackage{graphicx}
\usepackage{textcomp}
\usepackage{xcolor}
\usepackage[printonlyused]{acronym}
\usepackage[none]{hyphenat}
\usepackage{orcidlink}
\usepackage{adjustbox}
\usepackage{etoolbox}
\usepackage[caption=false,font=footnotesize,name=Fig.]{subfig}
\usepackage[T1]{fontenc}
\usepackage{fix-cm}
\usepackage{svg}
\usepackage{bm}
\usepackage{mathtools}
\def\BibTeX{{\rm B\kern-.05em{\sc i\kern-.025em b}\kern-.08em
    T\kern-.1667em\lower.7ex\hbox{E}\kern-.125emX}}

\renewcommand{\figurename}{Fig.}
\renewcommand{\thetable}{\Roman{table}}

\newcounter{algorithm}
\renewcommand{\thealgorithm}{\arabic{algorithm}}
\newcommand{\algcaption}[1]{%
    \refstepcounter{algorithm}%
    \textbf{Algorithm \thealgorithm.} #1%
}

  \acrodef{3G}{third generation}
  \acrodef{4G}{fourth generation}
  \acrodef{5G}{fifth generation}
  \acrodef{6G}{sixth generation}
  \acrodef{3GPP}{3rd generation partnership project}
  \acrodef{AI}{artificial intelligence}
  \acrodef{ADC}{analog-to-digital converter}
  \acrodef{AMF}{access and mobility function}
  \acrodef{AoA}{angle of arrival}
  \acrodef{AoD}{angle of departure}
  \acrodef{AR}{augmented reality}
  \acrodef{ASIC}{application-specific integrated circuit}
  \acrodef{AWGN}{additive white Gaussian noise}
  \acrodef{BER}{bit error rate}
  \acrodef{BFN}{beamforming network}
  \acrodef{BO}{back-off}
  \acrodef{BW}{bandwidth}
  \acrodef{C/A}{civilian acquisition}
  \acrodef{CA}{cell average}
  \acrodef{CCDF}{complementary cumulative density function}
  \acrodef{CFAR}{constant false alarm rate}
  \acrodef{CFO}{carrier frequency offset}
  \acrodef{CMMB}{China mobile multimedia broadcasting}
  \acrodef{CP}{cyclix prefix}
  \acrodef{CRLB}{Cramer-Rao lower bound}
  \acrodef{CSS}{chirp spread spectrum}
  \acrodef{CU}{centralized unit}
  \acrodef{DAC}{digital to analog converter}
  \acrodef{DBF}{digital beamforming}
  \acrodef{DFT}{discrete Fourier transform}
  \acrodef{DL-AoD}{downlink angle of departure}
  \acrodef{DLL}{delay locked loop}
  \acrodef{DL-OTDoA}{downlink observed time difference of arrival}
  \acrodef{DMRS}{demodulation reference signal}
  \acrodef{DSSS}{direct-sequence spread spectrum}
  \acrodef{DU}{distributed unit}
  \acrodef{DVB}{digital video broadcasting}
  \acrodef{E-CID}{enhanced cell id}
  \acrodef{ECDF}{empirical cumulative density function}
  \acrodef{E-SMLC}{evolved serving mobile location center}
  \acrodef{EIRP}{equivalent isotropic radiated power}
  \acrodef{EKF}{extended kalman filter}
  \acrodef{eNB}{evolved nodeb}
  \acrodef{EVM}{error vector magnitude}
  \acrodef{FHSS}{frequency hopping spread spectrum}
  \acrodef{FLL}{frequency locked loop}
  \acrodef{FFT}{fast Fourier transform}
  \acrodef{FFR}{full frequency re-use}
  \acrodef{FOV}{field of view}
  \acrodef{FSPL}{free space path loss}
  \acrodef{FR}{frequency region}
  \acrodef{FR1}{frequency region 1}
  \acrodef{FRF3}{frequency re-use factor 3}
  \acrodef{GDOP}{geometrical dilution of precision}
  \acrodef{GPS}{global positioning system}
  \acrodef{gNB}{next generation base station}
  \acrodef{GNSS}{global navigation satellite system}
  \acrodef{GS}{ground station}
  \acrodef{HAPS}{high-altitude platform systems}
  \acrodef{HIL}{hardware-in-the-loop}
  \acrodef{HPA}{high power amplifier}
  \acrodef{IBO}{input back-off}
  \acrodef{ICAL}{integrated communications and localization}
  \acrodef{IDFT}{inverse discrete Fourier transform}
  \acrodef{IFFT}{inverse fast Fourier transform}
  \acrodef{IoT}{internet of things}
  \acrodef{IIoT}{industrial internet of things}
  \acrodef{ICI}{inter-carrier interference}
  \acrodef{IMU}{inertial measurement unit}
  \acrodef{ISI}{inter-symbol interference}
  \acrodef{JCAP}{joint communication and positioning}
  \acrodef{KPI}{key performance indicator}
  \acrodef{LBA}{link budget analysis}
  \acrodef{LCS}{location-based services}
  \acrodef{LEO}{low earth orbit}
  \acrodef{LMC}{location management component}
  \acrodef{LMF}{location management function}
  \acrodef{LMMSE}{linear minimum mean squared error}
  \acrodef{LNA}{low noise amplifier}
  \acrodef{LOS}{line of sight}
  \acrodef{LPP}{localization positioning protocol}
  \acrodef{LPPa}{localization positioning protocol annex}
  \acrodef{LTE}{long term evolution}
  \acrodef{MCRB}{modified Cramer Rao bound}
  \acrodef{ML}{machine learning}
  \acrodef{MTA}{mean acquisition time}
  \acrodef{Multi-RTT}{multi-cell round trip time}
  \acrodef{NF}{network function}
  \acrodef{NSGA-II}{non-dominated sorting genetic algorithm II}
  \acrodef{NGSO}{non geostationary satellite orbit}
  \acrodef{NLOS}{non-line of sight}
  \acrodef{NMSE}{normalized mean square error}
  \acrodef{NR}{new radio}
  \acrodef{NTN}{non-terrestrial network}
  \acrodef{OFDM}{orthogonal frequency-division multiplexing}
  \acrodef{OMP}{orthogonal matching pursuit}
  \acrodef{OTA}{over-the-air}
  \acrodef{OTDoA}{observed time differential of arrival}
  \acrodef{OTFS}{orthogonal time frequency space}
  \acrodef{PAPR}{peak-to-average power ratio}
  \acrodef{PBCH}{physical broadcast channel}
  \acrodef{PDSCH}{physical downlink shared channel}
  \acrodef{PLL}{phase-locked loop}
  \acrodef{PNT}{positioning, navigation, and timing}
  \acrodef{POD}{precise orbit determination}
  \acrodef{PPP}{precise point positioning}
  \acrodef{PRN}{pseudo-random noise}
  \acrodef{PRS}{positioning reference signal}
  \acrodef{PSD}{power spectral density}
  \acrodef{PSS}{primary synchronization signal}
  \acrodef{PVT}{position, velocity and timing}
  \acrodef{QoS}{quality of service}
  \acrodef{RAN}{radio access network}
  \acrodef{RAT}{radio-access-technology}
  \acrodef{RB}{resource block}
  \acrodef{RE}{resource element}
  \acrodef{RG}{resource grid}
  \acrodef{RedCap}{reduced capacity}
  \acrodef{RMSE}{root mean square error}
  \acrodef{ROC}{receiver operating characteristic}
  \acrodef{RTK}{real time kinematics}
  \acrodef{RRC}{radio resource control}
  \acrodef{SBAS}{satellite based augmentation system}
  \acrodef{SDR}{software defined radio}
  \acrodef{SIB}{system information block}
  \acrodef{SIC}{sequential interference cancellation}
  \acrodef{SIR}{signal-to-interference ratio}
  \acrodef{SINR}{signal-to-interference plus noise ratio}
  \acrodef{SFN}{single frequency network}
  \acrodef{SLA}{service level agreement}
  \acrodef{SNR}{signal-to-noise ratio}
  \acrodef{SoO}{signal of opportunity}
  \acrodef{SoP}{signal of opportunity}
  \acrodef{SRS}{sounding reference signal}
  \acrodef{SRRC}{square root raised cosine}
  \acrodef{SS}{synchronization signal}
  \acrodef{SSB}{synchronization signal block}
  \acrodef{SSP}{subsatellite point}
  \acrodef{SSS}{secondary synchronization signal}
  \acrodef{TA}{timing advance}
  \acrodef{TC}{time coded}
  \acrodef{TC-OFDM}{time-coded orthogonal frequency division multiplexing}
  \acrodef{TDL}{tapped delay line}
  \acrodef{TN}{terrestrial network}
  \acrodef{ToA}{time of arrival}
  \acrodef{ToF}{time of flight}
  \acrodef{TS}{technical specification}
  \acrodef{TR}{technical report}
  \acrodef{UAV}{unmanned aerial vehicle}
  \acrodef{UE}{user equipment}
  \acrodef{UL-AoA}{uplink angle of arrival}
  \acrodef{UL-TDoA}{uplink time difference of arrival}
  \acrodef{UPA}{uniform planar array}
  \acrodef{VR}{virtual reality}
  \acrodef{WLAN}{wireless local area network}
  \acrodef{ZOH}{zero-order hold}

\begin{document}

\title{Fast 5G Signal Acquisition by Using Non-Uniform Sampling}

\author{Alejandro~Gonzalez-Garrido~\orcidlink{0000-0002-4695-8797}
 European Commission, Joint Research Centre (JRC), Ispra, Italy
\thanks{
Corresponding author: Alejandro Gonzalez Garrido, e-mail: alejandro.gonzalez-garrido@ec.europa.eu. Both authors contributed equally.
}\\
Carla Amatetti~\orcidlink{0000-0003-0253-2871},
University of Bologna}

\markboth{Transactions on Wireless Communications,~Vol.~XX, No.~XX, XX~202X}%
{Fast 5G Signal Acquisition by Using Non-Uniform Sampling}

\maketitle

\begin{abstract}
This paper proposes a framework for fast signal acquisition based on deterministic non-uniform sampling, with emphasis on multi-coset architectures and receivers driven by known synchronization sequences, pilots, or preambles. Unlike conventional sampling theory, which is formulated from a waveform-reconstruction perspective, the proposed approach is derived from the observation that acquisition is fundamentally a parametric inference problem in delay-Doppler space. Accordingly, the objective is not to reconstruct the full Nyquist-rate signal, but to preserve the statistics required for detection and estimation. The paper formulates compressed-domain acquisition through a generalized likelihood ratio test and shows how multi-coset sampling leads to reduced correlator structures operating directly on the retained samples. An offline exhaustive design procedure is introduced to select the coset pattern for a given sampling ratio by minimizing a cost that jointly enforces peak isolation in the acquisition surface and uniform retained-energy coverage over the delay search interval. The framework is evaluated on 5G NR synchronization using the PSS/SSS signals under a worst-case Doppler scenario. Results show that substantial reductions in mean acquisition time can be achieved relative to uniform sampling, with measured gains ranging from $2.8\times$ to $34.2\times$, depending on the selected compression ratio. The corresponding delay and Doppler root-mean-square errors quantify the estimation penalty introduced by aggressive sample reduction. These results demonstrate a clear complexity-performance trade-off and confirm the potential of multi-coset sampling for fast synchronization-oriented receivers.
\end{abstract}


\section{Introduction}
The conventional design of digital receivers is deeply rooted in the Nyquist--Shannon sampling theorem, according to which an arbitrary band-limited waveform can be represented without loss from uniform samples acquired at a rate not smaller than twice its highest frequency component \cite{nyquist_certain_1928,shannon_communication_1949}. This principle has shaped the design of \acp{ADC}, analog front-ends, and digital processing chains across communication, navigation, radar, and sensing systems. In its classical form, however, the sampling problem is commonly posed from a reconstruction perspective: the samples are acquired so that the waveform $x(t)$ itself can be faithfully recovered and then processed.
%\cite{nyquist_certain_1928,shannon_communication_1949}

In acquisition and synchronization, the receiver objective is more specific. The signal of interest is typically known up to a small set of unknown parameters, such as propagation delay, Doppler or carrier-frequency offset, complex amplitude, and, in spread-spectrum or pilot-aided systems, a code or preamble index. Accordingly, the task is not to reconstruct the entire received waveform, but rather to preserve the sufficient information required to decide whether the signal is present and to identify the location of the corresponding peak in the delay--Doppler--code search space. This inference-oriented viewpoint is explicit in low-rate time-delay estimation methods, compressive acquisition schemes for asynchronous multi-access signals, and compressive GPS acquisition architectures, all of which show that reliable synchronization can be carried out directly from reduced-dimensional measurements without first recovering the full Nyquist-rate signal \cite{gedalyahu_time-delay_2010,li_optimal_2012,li_gps_2012}. A closely related perspective also appears in sub-Nyquist radar, where target delay and Doppler are recovered from compressed measurements by operating directly on the parameters of interest rather than on a reconstructed high-rate waveform \cite{bar-ilan_sub-nyquist_2014}.

Multi-coset sampling is a deterministic non-uniform sampling architecture that, instead of retaining arbitrary irregular samples, selects a fixed subset of Nyquist instants within each block of $L$ samples, thereby producing a periodic non-uniform pattern that admits practical hardware implementation through synchronized low-rate branches \cite{venkataramani_perfect_2000}. At the same time, its algebraic structure is well suited to parametric inference, since the resulting measurements preserve enough information to support delay, frequency, and support estimation while avoiding the full-rate front-end demanded by exhaustive matched filtering \cite{venkataramani_perfect_2000,mishali_blind_2009,gedalyahu_time-delay_2010}. It can be especially relevant to acquisition and synchronization problems, so that receivers driven by known pilots, synchronization sequences, or spreading codes, make multi-coset sampling a natural candidate for reducing sampling and correlator complexity while maintaining acquisition capability \cite{li_optimal_2012,li_gps_2012}.

The objective of this paper is to present a theoretical framework for fast signal acquisition based on multi-coset sampling when the waveform contains known pilots. The emphasis is placed on acquisition and parameter estimation rather than on generic signal reconstruction. The main contributions are threefold. First, the acquisition problem is reformulated as compressed-domain detection and estimation using the generalized likelihood ratio test over a known pilot family. Second, multi-coset sampling is linked to reduced correlator structures that operate directly on the retained samples. Third, the selection of the sampling pattern is recast as a receiver optimization problem based on preservation of the delay-Doppler estimation.

\section{Signal Model and Problem Formulation}
\subsection{Acquisition as a Parametric Inference Problem}
Consider a received continuous-time signal containing a known pilot $s(t)$ affected by an unknown delay $\tau$, Doppler shift $\nu$, complex amplitude $\alpha$, and additive noise $w(t)$:
\begin{equation}
 r(t)=\alpha\, s(t-\tau)e^{j2\pi \nu t}+w(t).
 \label{eq:ct_model}
\end{equation}

After uniform Nyquist-rate sampling with period $T_s$, the discrete-time model becomes
\begin{equation}
 r[n]=\alpha\, s[n-d]e^{j2\pi \nu nT_s}+w[n],
 \label{eq:dt_model}
\end{equation}
with $d=\lfloor\tau / T_s \rfloor$ the discrete delay.

For a single path in white Gaussian noise, the maximum-likelihood acquisition problem reduces to a normalized matched-filter search over the unknown parameters. By eliminating the nuisance parameter $\alpha$, the estimator can be written as
\begin{equation}
    (\hat d,\hat\nu)=\arg\max_{d,\nu}\frac{\left|\langle r, s_{d,\nu}\rangle\right|^2}{\|s_{d,\nu}\|_2^2},\qquad s_{d,\nu}[n]=s[n-d]e^{j2\pi \nu nT_s}.
    \label{eq:ml_estimator_full}
\end{equation}

Hence, the key object for acquisition is the ambiguity surface $\chi(d,\nu)=\langle r, s_{d,\nu}\rangle,$ not the full waveform itself. This distinction is decisive: once the acquisition problem is recognized as an inference problem over a structured pilot family, sampling architectures may be optimized to preserve \eqref{eq:ml_estimator_full} rather than to reconstruct all Nyquist-rate samples.

\subsection{General Linear Measurement Model}
Let $\mathbf{x}\in\mathbb{C}^{N}$ denote the Nyquist-rate observation over a finite processing block. A general non-uniform sampling front-end is represented by the linear model
\begin{equation}
    \mathbf{y}=\mathbf{P}_{\Omega}\mathbf{x}.
    \label{eq:row_selection_model}
\end{equation}
where $\mathbf{P}_{\Omega}$ is a row-selection matrix associated with the retained index set $\Omega\subset\{0,\dots,N-1\}$. 

This formulation is standard in compressive signal processing and provides a clean bridge between classical matched filtering and reduced-dimensional inference \cite{davenport_signal_2010}.

The challenge is then to determine a structured sample set $\Omega$, that minimizes the processing burden while preserving the acquisition surface with sufficient fidelity. The present work focuses on multi-coset sampling because it yields a deterministic and implementation-friendly class of measurement operators.

\section{Multi-Coset Sampling and Compressed-Domain Acquisition}
\subsection{Multi-Coset Sampling Architecture}
Let the Nyquist period be $T_s$ and the corresponding Nyquist-rate sampling frequency be $f_s=1/T$. In multi-coset sampling, the receiver selects a period $L$, retains $K<L$ sample positions per period, and defines a coset set
\begin{equation}
    \mathcal{C}=\{c_0,c_1,\dots,c_{K-1}\}\subset\{0,1,\dots,L-1\}.
    \label{eq:coset_set}
\end{equation}

The retained samples occur at multiple times of the sampling period $T_s$
\begin{equation}
    t_i[m]=(mL+c_i)T_s, \qquad i=0,\dots,K-1,\; m\in\mathbb{Z}.
    \label{eq:multicoset_times}
\end{equation}

%As a visual example, Fig.~\ref{fig:NUS_example} represents an example of a random signal sampled uniformly and non uniformly.

%\begin{figure}
%    \centering
%    \includegraphics[width=1\linewidth]{Figures/sampling_visual_example.eps}
%    \caption{Example of a uniform sampling and non-uniform sampling}
%    \label{fig:NUS_example}
%\end{figure}

Equivalently, within each block of $L$ Nyquist instants, only $K$ samples are kept. The average sampling rate is therefore $f_{\mathrm{avg}}=\frac{K}{LT_s}.$ This interpretation immediately exposes the rate-reduction factor and clarifies why the ratio $K/L$ is the main design variable. Multi-coset sampling can also be viewed as a bank of $K$ uniform low-rate sequences, each associated with one retained coset, which makes the architecture attractive for digital realization and branch-based processing \cite{mishali_blind_2009}.

\subsection{Compressed-Domain Detection and Estimation}
Assume now that the block observation belongs to the parametric family
\begin{equation}
    \mathbf{x}=\alpha\,\mathbf{s}(\theta)+\mathbf{w},\qquad \theta=(d,\nu,\ldots),
    \label{eq:parametric_family}
\end{equation}
where $\mathbf{s}(\theta)$ is the Nyquist-rate local copy of the pilots under hypothesis $\theta$. After multi-coset sampling,
\begin{equation}
    \mathbf{y}=\mathbf{P}_{\Omega}\mathbf{x}=\alpha\,\mathbf{P}_{\Omega}\mathbf{s}(\theta)+\mathbf{P}_{\Omega}\mathbf{w}.
    \label{eq:compressed_parametric_family}
\end{equation}

Defining the compressed pilot $\mathbf{a}(\theta)=\mathbf{P}_{\Omega}\mathbf{s}(\theta),$ the generalized likelihood ratio test in white Gaussian noise is
\begin{equation}
    \Lambda_{\Omega}(\theta)=\frac{\left|\mathbf{a}(\theta)^H\mathbf{y}\right|^2}{\|\mathbf{a}(\theta)\|_2^2},
    \label{eq:compressed_glrt}
\end{equation}

and the corresponding estimator is
\begin{equation}
    \hat\theta=\arg\max_{\theta\in\Theta}\Lambda_{\Omega}(\theta).
    \label{eq:compressed_estimator}
\end{equation}

%When the sampled noise is colored with covariance $\mathbf{R}_w$, the correct whitened form becomes
%\begin{equation}
%    \Lambda_{\Omega}(\theta)= \frac{\left|\mathbf{a}(\theta)^H\mathbf{R}_w^{-1}\mathbf{y}\right|^2}{\mathbf{a}(\theta)^H\mathbf{R}_w^{-1}\mathbf{a}(\theta)},
%    \label{eq:compressed_glrt_colored}
%\end{equation}
%where $\mathbf{R}_w=\mathbb{E}[\mathbf{P}_{\Omega}\mathbf{w}\mathbf{w}^{H}\mathbf{P}_{\Omega}^{H}]$.

Equation~\eqref{eq:compressed_glrt}  formalize the central principle of task-aware sampling: the receiver may operate directly on the reduced samples without first reconstructing the Nyquist-rate waveform, provided that the compressed pilots retain sufficient discriminability over the hypothesis space \cite{davenport_signal_2010}.

\section{Coset Design for Acquisition and Parameter Estimation}
The coset pattern adopted in the numerical evaluation is obtained by an \emph{offline exhaustive search} that is directly matched to the reduced correlator used later during acquisition and parameter estimation. For a fixed pair $(L,K)$, all $\binom{L}{K}$ admissible multi-coset patterns are enumerated and scored with the same normalized delay-Doppler acquisition metric implemented by the receiver. Consequently, the selected pattern is not obtained from a generic reconstruction-oriented criterion, but from the actual behaviour of the reduced correlator on the synchronization task of interest.

The design objective is twofold. First, the reduced acquisition surface must have a dominant peak at the correct delay-Doppler hypothesis and low off-peak responses elsewhere. Second, the retained pilot energy should remain as uniform as possible over the searched delay interval, so that the reduced correlator does not become biased toward a subset of delays simply because they are better covered by the retained samples. These two objectives are jointly enforced by the cost function described below.

\subsection{Design Scenario and Candidate Patterns}
For a fixed period $L$ and number of retained cosets $K$, let
\begin{equation}
    \mathcal{C}=\{c_0,c_1,\ldots,c_{K-1}\}\subset\{0,1,\ldots,L-1\}
    \label{eq:coset_set_design}
\end{equation}
be one candidate pattern. The corresponding binary sampling mask over an observation window of length $N_{\mathrm{obs}}$ is defined as
\begin{equation}
    m_{\mathcal{C}}[n]=\begin{cases}
                        1, & n \bmod L \in \mathcal{C},\\
                        0, & \text{otherwise}.
                        \end{cases}
    \label{eq:coset_mask_design}
\end{equation}
In the implementation, the observation length is chosen as  $N_{\mathrm{obs}} = N_s + d_{\max}$, where $N_s$ is the pilot length and $d_{\max}$ is the maximum delay searched by the receiver.

To evaluate each candidate pattern, the algorithm uses a small set of representative true delays,
\begin{equation}
    \mathcal{D}_{\mathrm{ref}}=
    \left\{
    \left\lfloor \frac{d_{\max}}{4} \right\rfloor,
    \left\lfloor \frac{d_{\max}}{2} \right\rfloor,
    \left\lfloor \frac{3d_{\max}}{4} \right\rfloor
    \right\},
    \label{eq:reference_delays}
\end{equation}
which are selected to probe the early, central, and late portions of the delay search interval. %The reference Doppler during the design stage is set to zero and associated with the nearest bin of the design Doppler grid. This choice is sufficient because the objective of the coset search is to preserve the peak geometry of the reduced ambiguity surface under the same search engine later used in the Monte Carlo evaluation.

\subsection{Reduced Acquisition Metric Used for the Design}
For each candidate pattern $\mathcal{C}$ and each reference delay $d_0\in\mathcal{D}_{\mathrm{ref}}$, the script synthesizes a noiseless received signal according to the same pilot model used in the acquisition stage,
\begin{equation}
    r_{d_0}[n]=s[n-d_0]e^{j2\pi\nu_{p} nT_s},
    \label{eq:noiseless_design_signal}
\end{equation}
where $\nu_{p}$ are the Doppler hypothesis to generate the coset. The candidate mask $m_{\mathcal{C}}[n]$ is then applied to this observation and the normalized reduced correlator is evaluated over the complete delay-Doppler search grid. For a test delay $d$ and Doppler $\nu$, the reduced correlation is
\begin{equation}
    \chi_{\mathcal{C}}(d,\nu)=\sum_{n=0}^{N_{\mathrm{obs}}-1} m_{\mathcal{C}}[n] \, r_{d_0}[n] \, s^*[n-d] \, e^{-j2\pi \nu nT_s}.
    \label{eq:design_reduced_corr}
\end{equation}
The corresponding normalized acquisition metric is
\begin{equation}
    \Lambda_{\mathcal{C}}(d,\nu)=\frac{|\chi_{\mathcal{C}}(d,\nu)|^2}{D_{\mathcal{C}}(d)},
    \label{eq:design_metric}
\end{equation}
with delay-dependent retained pilot energy
\begin{equation}
    D_{\mathcal{C}}(d)=\sum_{n=0}^{N_{\mathrm{obs}}-1} m_{\mathcal{C}}[n] \, |s[n-d]|^2.
    \label{eq:design_denom}
\end{equation}
The normalization in \eqref{eq:design_metric} is important because different coset patterns retain different fractions of the pilot energy as the delay varies. Without this normalization, a pattern could be favoured only because it overlaps more strongly with a subset of delays, rather than because it produces a better acquisition surface.

\subsection{Peak-Isolation Score and Energy-Balance Regularization}
The first component of the design cost measures how well the dominant acquisition peak is isolated from the remaining hypotheses. Let $i_0$ denote the Doppler-bin index for zero Doppler, this has been extended to other Doppler hypotheses in Algorithm~\ref{alg:coset}. For a given reference delay $d_0$, the algorithm defines a guard region around the true peak,
\begin{equation}
    \mathcal{G}(d_0,i_0)=\left\{(d,i): |d-d_0|\leq 1,\ |i-i_0|\leq 1\right\},
    \label{eq:guard_region}
\end{equation}
and excludes that neighborhood when searching for the largest sidelobe. The resulting normalized sidelobe-to-peak ratio is
\begin{equation}
    \mathrm{SPR}_{\mathcal{C}}(d_0)=\frac{\max\limits_{(d,i)\notin \mathcal{G}(d_0,i_0)}\Lambda_{\mathcal{C}}(d,\nu_i)}{\Lambda_{\mathcal{C}}(d_0,\nu_{i_0})},
    \label{eq:spr_cost}
\end{equation}
where $f_i$ denotes the $i$th Doppler bin of the design grid. Lower values of \eqref{eq:spr_cost} indicate a more pronounced and less ambiguous main peak.

The second component of the design cost measures how uniformly the retained samples cover the pilot energy over the delay search. For the same candidate pattern, the script computes the energy-balance term
\begin{equation}
    B_{\mathcal{C}}(d)=\frac{\min\limits_{d} D_{\mathcal{C}}(d)}{\max\limits_{d} D_{\mathcal{C}}(d)},\qquad 0 < B_{\mathcal{C}}(d) \leq 1.
    \label{eq:energy_balance}
\end{equation}
A value close to one indicates that the reduced correlator retains a nearly uniform fraction of the pilot energy over all delays, whereas a small value indicates strong delay-dependent coverage.

The final cost assigned to a candidate pattern is obtained by averaging the two terms over the reference delays and penalizing poor energy balance,
\begin{equation}
    J(\mathcal{C})=\frac{\overline{\mathrm{SPR}}_{\mathcal{C}}}{\overline{B}_{\mathcal{C}}},
    \label{eq:final_cost}
\end{equation}
where
\begin{equation}
    \overline{\mathrm{SPR}}_{\mathcal{C}}=\frac{1}{|\mathcal{D}_{\mathrm{ref}}|}\sum_{d_0\in\mathcal{D}_{\mathrm{ref}}}\mathrm{SPR}_{\mathcal{C}}(d_0)
    \label{eq:avg_pslr}
\end{equation}
and
\begin{equation}
    \overline{B}_{\mathcal{C}}=\frac{1}{|\mathcal{D}_{\mathrm{ref}}|}\sum_{d_0\in\mathcal{D}_{\mathrm{ref}}}B_{\mathcal{C}}(d_0).
    \label{eq:avg_balance}
\end{equation}
Therefore, the preferred pattern is the one that minimizes the average normalized sidelobe level while maintaining a well-balanced retained-energy profile across the delay search window.

\subsection{Exhaustive Search Procedure}
For fixed $(L,K)$, the implementation evaluates all admissible coset sets and selects the one that minimizes \eqref{eq:final_cost}. The procedure is summarized in Algorithm~\ref{alg:coset}.

\begin{figure}[!t]
\centering
\begin{minipage}{0.98\columnwidth}
\hrule
\vspace{0.5ex}
\algcaption{Offline coset selection for a fixed pair $(L,K)$.}\label{alg:coset}
\vspace{0.5ex}
\hrule
\vspace{0.5ex}

\begin{algorithmic}[1]
\STATE \textbf{Input:} pilot $s[n]$, maximum delay $d_{\max}$, design Doppler grid $\mathcal{F}_{\mathrm{des}}$, period $L$, number of kept cosets $K$
\STATE Form the candidate set of all $\binom{L}{K}$ coset patterns
\STATE Define the reference delay set $\mathcal{D}_{\mathrm{ref}}$ according to \eqref{eq:reference_delays}
\FOR{each candidate pattern $\mathcal{C}$}
    \STATE Build the binary mask $m_{\mathcal{C}}[n]$ from \eqref{eq:coset_mask_design}
    \FOR{each reference delay $d_0 \in \mathcal{D}_{\mathrm{ref}}$}
        \FOR{each reference Doppler $\nu \in \mathcal{F}_{\mathrm{des}}$}
            \STATE Generate the noiseless delayed-Doppler signal $r_{d_0}[n]$
            \STATE Compute the reduced acquisition surface $\Lambda_{\mathcal{C}}(d,\nu)$ over the full design grid
            \STATE Evaluate $\mathrm{SPR}_{\mathcal{C}}(d_0)$ using \eqref{eq:spr_cost}
            \STATE Evaluate $B_{\mathcal{C}}(d)$ using \eqref{eq:energy_balance}
        \ENDFOR
    \ENDFOR
    \STATE Compute the aggregate cost $J(\mathcal{C})$ in \eqref{eq:final_cost}
\ENDFOR
\STATE \textbf{Output:} $\mathcal{C}^{\star} = \arg\min_{\mathcal{C}} J(\mathcal{C})$
\end{algorithmic}

\vspace{0.5ex}
\hrule
\end{minipage}
\end{figure}

Because the search is exhaustive, this procedure is computationally feasible only for moderate values of $L$ and $K$. This is sufficient for the present study, where the objective is to characterize the complexity-performance trade-off of reduced correlators rather than to solve a very large-scale combinatorial design problem.

\subsection{Role of the Period and Keep Parameters}
Algorithm~\ref{alg:coset} separates the choice of the coset set from the choice of the pair $(L,K)$. For any fixed $(L,K)$, Algorithm~\ref{alg:coset} returns the best pattern $\mathcal{C}^{\star}$ among all admissible combinations. The global choice of $L$ and $K$ is then performed empirically by repeating the full simulation for several pairs and comparing the resulting acquisition time, detection performance, and delay/Doppler estimation errors. Accordingly, the overall methodology is two-level: an inner exhaustive search that determines the best coset pattern for a fixed compression structure; and an outer performance-complexity evaluation that determines which compression ratio is acceptable for the receiver under study.

\subsection{Suitability for Preambles, Pilots, and Known Synchronization Signals}
Known preambles and pilots are particularly suitable for non-uniform sampling because the receiver faces a low-dimensional inference problem embedded in a high-dimensional waveform space. For a typical synchronization stage, the unknown quantities are restricted to a small set, often including a delay, a Doppler or residual carrier-frequency offset, one or a few complex channel coefficients, and possibly a discrete hypothesis index identifying a codeword or preamble. This structure explains why pilot-based acquisition can tolerate significantly more aggressive measurement reduction than arbitrary data demodulation, where the unknown payload symbols themselves occupy a large space dimension.

For example, in \ac{NR} an \ac{SS}/\ac{PBCH} block spans four \ac{OFDM} symbols over 240 contiguous subcarriers, while the \ac{PSS} and \ac{SSS} occupy only symbols $0$ and $2$ and are mapped to subcarriers $56$-$182$ \cite{3gpp_36211_2021_4}. This structure suggests a hierarchical receiver strategy: first restrict acquisition to the burst intervals and symbols that can contain synchronization energy, and only then apply multi-coset sampling and compressed-domain processing within those windows. The same design philosophy extends to other \ac{OFDM} preambles and pilot bursts.

\section{Delay-Doppler Estimation and Reduced Correlator Structures}
\subsection{Single-pilot Formulation}
Let $s[n]$, $0\le n<N_s$, denote the Nyquist-rate discrete-time pilot. The received signal under a single-path hypothesis is
\begin{equation}
    r[n]=\alpha\,s[n-d]e^{j2\pi \nu nT_s}+w[n].
    \label{eq:single_pilot_model}
\end{equation}
The full-rate delay-Doppler correlator is
\begin{equation}
    \chi(d,\nu)=\sum_n r[n]s^*[n-d]e^{-j2\pi \nu nT_s}.
    \label{eq:full_correlator}
\end{equation}
Under multi-coset sampling, only the indices in $\Omega$ are available, so the reduced correlator becomes
\begin{equation}
    \chi_{\Omega}(d,\nu)=\sum_{n\in\Omega} r[n]s^*[n-d]e^{-j2\pi \nu nT_s}.
    \label{eq:reduced_correlator}
\end{equation}
The corresponding normalized acquisition metric is
\begin{equation}
    \Lambda_{\Omega}(d,\nu)=\frac{|\chi_{\Omega}(d,\nu)|^2}{\sum_{n\in\Omega}|s[n-d]|^2}.
    \label{eq:normalized_reduced_metric}
\end{equation}
The theoretical requirement is therefore transparent: the coset pattern must preserve the location, sharpness, and contrast of the dominant peak of the ambiguity surface under compression.

\subsection{Multiple Pilots, Preambles, or Code Hypotheses}
Many practical acquisition problems involve a discrete set of candidate sequences or codewords. If the receiver must jointly search over a family $\{s_m\}$ of known sequences indexed by a discrete hypothesis variable $m$ and a parameter vector $\theta$, define the compressed atom
\begin{equation}
    \mathbf{a}_{m,\theta}=\mathbf{P}_{\Omega}\mathbf{s}_m(\theta).
    \label{eq:compressed_atom_family}
\end{equation}
The joint acquisition problem becomes
\begin{equation}
    (\hat m,\hat\theta)=\arg\max_{m,\theta}
    \frac{\left|\mathbf{a}_{m,\theta}^H\mathbf{y}\right|^2}{\|\mathbf{a}_{m,\theta}\|_2^2}.
    \label{eq:joint_code_param_search}
\end{equation}
This model directly captures synchronization based on \ac{DSSS} spreading codes, \ac{OFDM} preambles, pilot bursts, and cell-identifier search in standards that embed known synchronization sequences.

\section{Results}
This section shows the results of using this non-uniform sampling to acquire the \ac{PSS}/\ac{SSS} pilots in 5G. The \acp{KPI} evaluated for this non-uniform sampling are the \ac{RMSE} of the delay and Doppler estimations as the non-uniform sampling is optimized for the estimation of these parameters. In addition, the processing gain in the acquisition stage is reported.

The parameters of the scenario are summarized in Table \ref{tab:scenario_parameters}. 

\begin{table}
    \centering
    \caption{Scenario parameters}
    \begin{tabular}{ccc}
        \toprule
        Parameter & Units & Value\\ 
        \midrule
        Doppler max & kHz & $\pm20$\\  
        Subcarrier spacing & kHz & 15 \\ 
        Monte Carlo trials & \# & 5000\\ 
        Sampling frequency & MHz & 3.84 \\ 
        $N_{\mathrm{FFT}}$ & \# & 256 \\ 
        \bottomrule
    \end{tabular}
    \label{tab:scenario_parameters}
\end{table}

One parameter not listed in Table \ref{tab:scenario_parameters} is the number of Doppler bins (hypothesis to test). A common assumption in \ac{GNSS} literature \cite{kaplan2006understanding}, is to allow a maximum loss of  $-3\,\mathrm{dB}$. Then, $\Delta f\approx\frac{0.443}{T_{\mathrm{obs}}}$, where $T_{\mathrm{obs}}$ is the observation time used to estimate the Doppler, in this case, we assume a 5G subframe containing the \ac{PSS} and \ac{SSS}. Therefore, the Doppler bin width in our scenario is $\Delta f\approx443\,\mathrm{Hz}$.

The first result is the comparison of the processing gain of using non-uniform sampling with respect to the classic uniform sampling. Table \ref{tab:processing_gain} shows the measured \ac{MTA} and the gain for different ratios of $L/K$ compared with the uniform sampling.

\begin{table}[t]
    \centering
    \caption{Mean acquisition time.}
    \label{tab:processing_gain}
    \begin{tabular}{cccc}
        \toprule
        Config & \makecell{Sample\\instant $\mathcal{C}^{\star}$} & \makecell{MTA\\ms} & \makecell{Gain}  \\
        \midrule
        All samples & -                           & $9.376$ & $1.0\times$ \\
        8/4       & $[2,3,4,5]$                     & $3.337$ & $2.8\times$  \\
        16/8      & \makecell{$[2,3,4,5,$\\$10,11,12,13]$} & $2.995$ & $3.1\times$  \\
        8/2       & $[3,5]$                         & $1.357$ & $6.9\times$  \\
        16/1      & $[12]$                          & $0.612$ & $15.3\times$ \\
        32/1      & $[12]$                          & $0.274$ & $34.2\times$ \\
        \bottomrule
    \end{tabular}
\end{table}

\begin{figure}
    \centering
    \includegraphics[width=1\linewidth]{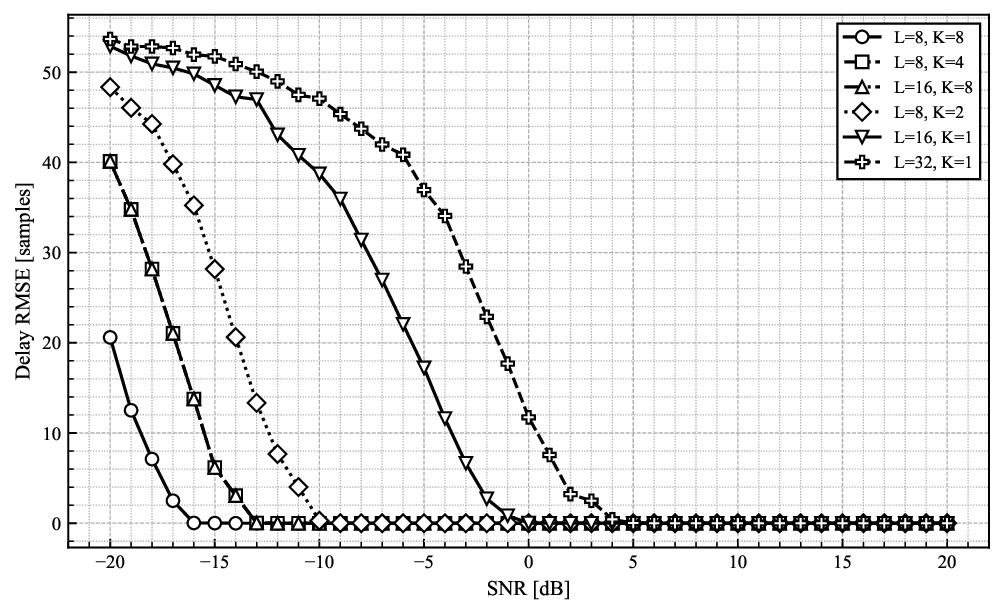}
    \caption{Delay estimation RMSE for different combinations of non uniform sampling schemes. Uniform sampling as benchmark.}
    \label{fig:delay_estimation}
\end{figure}

Then, in order to see the impact of this reduced number of samples has in the delay estimation, Fig.~\ref{fig:delay_estimation} shows the delay \ac{RMSE} obtained under different $L/K$ ratios and compares it to the uniform-sampling benchmark.

\begin{figure}
    \centering
    \includegraphics[width=1\linewidth]{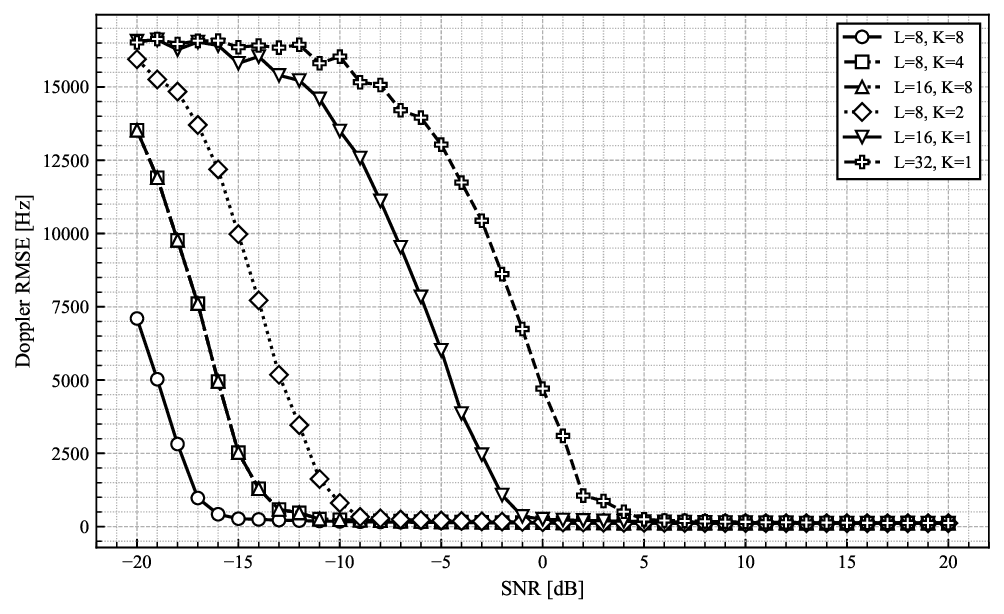}
    \caption{Doppler estimation RMSE for different combinations of non uniform sampling schemes. Uniform sampling as benchmark.}
    \label{fig:doppler_estimation}
\end{figure}

Finally, Fig.~\ref{fig:doppler_estimation} shows the Doppler \ac{RMSE} of using a classic uniform sampling versus different ratios of $L/K$.

These results show the trade-off to pay when going to a non-uniform sampling versus a classical sampling. It is worth mentioning that depending on the application, and the next steps on the receiver, one can assume the extra error of using non-uniform sampling for getting a huge benefit in processing time.

\section{Conclusion}
This paper presents a framework for fast acquisition based on deterministic non-uniform sampling, driven by known preambles, pilots, or spreading codes. The main result is that the appropriate design objective for synchronization-oriented receivers is not waveform reconstruction, but the preservation of the acquisition and estimation statistics defined over the delay-Doppler hypothesis space. Under this viewpoint, multi-coset sampling emerges as a solution because it combines a deterministic implementation with a reduced correlator structure.

From a practical perspective, the framework is especially relevant for synchronization tasks, where the waveform is known and the number of unknown parameters is small compared to the  signal space. In such cases, a well-designed multi-coset front-end can provide a meaningful reduction in the acquisition time with little degradation on the quality of the estimates. Future work should address the effect of multipath, and experimental validation with \ac{5G} signals.

\bibliographystyle{IEEEtran}
\bibliography{references}

\end{document}